\documentstyle[12pt,epsfig]{article}

\def\beq{\begin{equation}}
\def\eeq{\end{equation}}
\def\bea{\begin{eqnarray}}
\def\eea{\end{eqnarray}}
\def\noi{\noindent}
\def\pnq{\psi_{\nu}(q)}
\def\pnqc{\psi_{\nu}^*(q)}
\def\ppnq{\psi_{\nu^\prime}(q)}

\begin{document}

\rightline{DESY 98-108}
\rightline{hep-ph/9808340}
\rightline{\today}
\vspace{1cm}

\begin{center}
{\Large\bf On the 2nd Order Corrections to the Hard Pomeron\\
and the Running Coupling}\vspace{1cm}

N. Armesto$^a$, J. Bartels$^a$ and M. A. Braun$^b$\vspace{0.5cm}

$^a$ {\it II. Institut f\"ur Theoretische Physik, Universit\"at
Hamburg,} \\
{\it Luruper Chaussee 149, D-22761 Hamburg, Germany} \\

\vspace{0.2cm}

$^b$ {\it Department of
High-Energy Physics, University of St. Petersburg,}\\
{\it 198904 St. Petersburg, Russia}\\

\end{center}

\vspace{2cm}
{\small
\centerline{\bf Abstract}
It is shown that solutions to the 2nd order BFKL eigenvalue
equation exist for
arbitrary large real
values  of the complex angular momentum $j$. This corresponds
to a cut in the complex $j$ plane along the whole real axis, and it makes
the use of the complex angular momentum variable for the calculation of
the high-energy behavior somewhat questionable. The eigenfunctions
contain non-perturbative pieces which behave as $\exp(-1/\alpha_s b)$ and
have no counterpart in the leading-log BFKL equation. The high-energy
behavior of the 2nd order BFKL Green function as found by other authors,
is reproduced by excluding these non-perturbative pieces of the
eigenfunctions.
}

\newpage
\section{Introduction}
Recently corrections of the 2nd order in the coupling constant
to the hard pomeron (BFKL) equation have been calculated \cite{1r}.
For the intercept they are
accompanied by a large negative coefficient \cite{1r,3r},
which makes its prediction
reliable only for extremely low coupling constants and translates into
extremely large relevant mass scales. Although
this conclusion has been questioned in \cite{4r}, different calculations
\cite{2r,levin,bv} show features of this approach which clearly limit
its applicability.

In \cite{2r} the corrections to the BFKL
kernel
have been summed to obtain the corrected high-energy behavior of the
corresponding amplitudes. This summation resulted  in a multiple
integral
over intermediate rapidities, which was performed  in the saddle point
approximation and using the asymptotic form of the 1s order (leading-log)
BFKL Green
function. Even after these simplifications this leads to a highly
nontrivial
multiple integral, the calculation of which required much skill and
ingenuity. A similar result for the high behavior has been found
also in ~\cite{levin}.

In this  note we draw attention to the fact that there exists a simple
way of exactly solving the BFKL eigenvalue
equation with the kernel being known in the two first
orders. The solution turns out to exist for any real value of
the complex angular momentum $j$, from $-\infty$ to $+\infty$.
It means that the singularities of the BFKL amplitude occupy the whole real
axis in the $j$ plane, so that there is no rightmost singularity and no
intercept (the first hint of a ``non-Regge" behavior was noted in \cite{2r}
upon the observation
of a non-usual form of the $s$-dependence). All solutions to the
eigenvalue equation turn out to have
non-perturbative features. They separate into
``normal" pieces, which can be related to
the lowest order solutions and behave similarly both in $q$ and the
energy,
and ``abnormal" contributions,
which are non-oscillatory and proportional to $\exp(-1/\alpha_s b)$.
If we  neglect the abnormal pieces, we obtain
an asymptotic behavior at high $s$, which coincides with \cite{2r}.

Our method easily generalizes to the case when the coupling constant is summed
to all orders to run with the momentum scale. In the lowest order in this
running coupling the equation still can be written, in spite of the evident
difficulties at small momenta. Solutions of this equation
also exist for any real $j$.
This indicates that the physical equation
should have a kernel which is different from the lowest order one, not only
in the running of the coupling but also in its functional dependence, to
give a meaningful result in terms of a Mellin transform.
When this paper was being completed, a preprint by E. M. Levin appeared
\cite{levin}, in
which a very similar method has been applied in order to solve the 2nd order 
BFKL equation, with the running coupling constant summed to all orders and 
being modified in a specific manner in the confinement region.
The contents of this preprint partly 
overlap with our Section 4. However the central point of our paper - the 
existence of a cut in the $j$ plane along the whole real axis  - is not 
discussed in \cite{levin}.

\section{Basic equations}

We start by  presenting basic formulas necessary for our derivation. They
all are either standard or can be taken from Refs. \cite{1r,2r}. We restrict
ourselves
to the forward case and to the azimuthally symmetric wave function
$\psi(q)$,
which is known to dominate in the high-energy limit.
The pomeron equation will be written in the form of a Schr\"odinger
equation
\beq
H\psi(q)=E\psi(q), \label{eq1}
\eeq
where the ``energy" is related to the complex angular momentum by $E=1-j
\ (=-\omega$ in the usual notation \cite{1r}).
The ``Hamiltonian" $H$ has been calculated up to terms of the second order in
the coupling constant $\alpha_s(\mu^2)\equiv\alpha_s$:
\beq
H=H^{(1)}+H^{(2)}. \label{eq2}
\eeq
Proper functions of the lowest order (BFKL) Hamiltonian $H^{(1)}$ are
well-known:
\beq
H\psi_{\nu}(q)=\epsilon^{(1)}_{\nu}\psi_{\nu}(q), \label{eq3}
\eeq
where
\beq
\pnq=\frac{1}{\pi q\sqrt{2}}q^{2i\nu}, \label{eq4}
\eeq
the unperturbed energies are (in units of $\alpha_sN_c/\pi$)
\beq
\epsilon^{(1)}_{\nu}=\psi\left(\frac{1}{2}+i\nu\right)+
\psi\left(\frac{1}{2}-i\nu\right)-2\psi(1) \label{eq5}
\eeq
($\epsilon^{(1)}_{\nu}=-\chi(\gamma)$, $\gamma=1/2+i\nu$, in the usual notation
\cite{1r}).
In this equation $\psi(z)=d\ln{\Gamma(z)}/dz$,
and $\nu$ runs from $-\infty$ to $+\infty$. The wave functions (\ref{eq4}) are
correctly normalized:
\beq
\langle \psi_{\nu}|\psi_{\nu^\prime}\rangle\equiv
\int d^2q\pnqc\ppnq =\delta(\nu-\nu^\prime). \label{eq6}
\eeq

The second order part of the Hamiltonian can be conveniently  written
in terms of its action on the proper functions of the first order
Hamiltonian $\psi_{\nu}$ \cite{1r}.
Namely
\beq
H\pnq=[\kappa_{sc}(\nu)+\kappa_r(\nu,q)]\pnq. \label{eq7}
\eeq
The first term  in the bracket is scale invariant:
\beq
\kappa_{sc}(\nu)=-\frac{\alpha_sN_c}{4\pi}c(\nu)\epsilon^{(1)}_{\nu},
\label{eq8}
\eeq
where the function $c(\nu)$ can be found in \cite{1r}.
The second term provides for the running of the coupling:
\beq
\kappa_r(\nu,q)=[-2\alpha_sb\,\ln (q/\mu)]\ \epsilon^{(1)}_{\nu},
\label{eq9}
\eeq
where $b=(11N_c-2N_f)/(12\pi)$.
We put $\mu=1$ for simplicity.

$\kappa_{sc}$ is independent of $q$. So it simply shifts the energy
\beq
\epsilon^{(1)}_{\nu}\rightarrow \epsilon_{\nu}=
\epsilon^{(1)}_{\nu}\left(1-\frac{\alpha_s
N_c}{4\pi}c(\nu)\right). \label{eq10}
\eeq
As a result the lowest energy level at $\nu=0$ acquires a factor
\[1-\frac{\alpha_sN_c}{4\pi}c(0)\simeq 1-\frac{\alpha_sN_c}{\pi}\, 6.562\]
for $N_c=3$ and $N_f=2$,
which is so disturbing from the point of practical application of this
formalism to present momentum scales.

The second term $\kappa_r$ depends on $q$ and evidently changes the
equation itself. In \cite{2r} the influence of this factor was investigated by
summing all orders in its action. We are going to propose a different
approach.

\section{The
$\nu$-representation}
Our idea is quite trivial (and not new, see \cite{levin}).
Since the action of the
second order Hamiltonian on the proper functions $\psi_{\nu}$ is known, we are
going to pass to the representation which these functions provide. To this aim
we
present any solution $\Psi(q)$ of the total Hamiltonian
$H$ as a superposition of $\psi_{\nu}$,
\beq
\Psi(q)=\int d\nu f(\nu)\pnq. \label{eq11}
\eeq
Mathematically it is nothing but a
Fourier transformation of $q\Psi(q)$ with respect to $\ln q$. Using
(\ref{eq6}) we have
\beq
f(\nu)=\int d^2q\pnqc\Psi(q). \label{eq12}
\eeq

In the $\nu$ representation the Schr\"odinger equation (\ref{eq1}) reads
\beq
\int d\nu^\prime H(\nu,\nu^\prime)f(\nu^\prime)=Ef(\nu), \label{eq13}
\eeq
where, of course,
\beq
H(\nu,\nu^\prime)=\langle\psi_{\nu}|H|\psi_{\nu^\prime}\rangle=
\int d^2q\pnqc H\ppnq. \label{eq14}
\eeq

Using (\ref{eq3}) we obtain
\beq
H^{(1)}(\nu,\nu^\prime)=\epsilon_{\nu}^{(1)}\delta(\nu-\nu^\prime).
\label{eq15}
\eeq
The scale invariant part in $H^{(2)}$ gives similarly
\beq
H^{(2)}_{sc}(\nu,\nu^\prime)=
-\frac{\alpha_cN_c}{4\pi}c(\nu)\epsilon_{\nu}^{(1)}\delta(\nu-\nu^\prime).
\label{eq16}
\eeq
The running coupling part involves $\ln q$, which leads to a derivative
of the $\delta$-function,
\beq
H^{(2)}_r(\nu,\nu^\prime)=
-i\alpha_s b\epsilon^{(1)}_{\nu^\prime}\delta^\prime(\nu-\nu^\prime).
\label{eq17}
\eeq
We note that in all second order terms
we can substitute $\epsilon^{(1)}_{\nu}$ by $\epsilon_{\nu}$ defined by
(\ref{eq10})  in the adopted approximation.

Thus the Schr\"odinger equation in the $\nu$ representation turns out to be
a differential equation of the first order in $\nu$:
\beq
(-i\alpha_s b)\frac{d}{d\nu}\left[
\epsilon_{\nu}f(\nu)\right]+\epsilon_{\nu}
f(\nu)=Ef(\nu). \label{eq18}
\eeq
This equation is solved trivially. The solution can be taken as
\beq
f_E(\nu)=\frac{C}{\epsilon_{\nu}}\exp\left(\frac{iE}{\alpha_s
b}\int^{\nu}_0\frac{d\nu^\prime}
{\epsilon_{\nu^\prime}}-\frac{i\nu}{\alpha_s b}\right), \label{eq19}
\eeq
where $C$ is a constant. Note that the 1st order
$\epsilon_{\nu}^{(1)}$ has zeros
at $\nu=\nu_0\simeq\pm 0.6375$. So, in this approximation, one has to specify
the rule to circumvent the pole singularity in the integral, say, change
$\epsilon_{\nu}\rightarrow\epsilon_{\nu}-i0$. At the 2nd order $\epsilon_{\nu}$
does not vanish for real $\nu$, so that this change is not necessary. However
in both cases $\epsilon_{\nu}$ contains an imaginary part. Correspondingly,
for the conjugate equation one has to take $\overline{\epsilon}_{\nu}=
\epsilon_{\nu}^*$. Then it will have similar solutions $\overline{f}_E$ as
(\ref{eq19})
without the prefactor
$1/\epsilon_{\nu}$ and with an opposite sign of ${\rm Im}\,\epsilon_{\nu}$.
Imaginary parts of $\epsilon_{\nu}$ and $\overline{\epsilon}_{\nu}$
will produce
certain real factors in $f_E$ and $\overline{f}_E$, of a
non-perturbative nature
(with $1/\alpha_s$ in the exponent), the product of which gives unity.
Because of that in a product $f_E\overline
{f}^*_E$, which is of practical importance,
these products always cancel. Due to this circumstance, in future we
shall neglect the real factors  generated, for real $\nu$, by the
imaginary part of $\epsilon_{\nu}$.

One can see that  $f_E$ and  $\overline{f}_E$ form an orthonormal system:
Indeed
\beq
\langle\overline{f}_E|f_{E^\prime}\rangle=|C|^2\int\frac{d\nu}{\epsilon_{\nu}}
\exp\left(\frac{i(E^\prime-E)z(\nu)}{\alpha_sb}\right), \label{eq20}
\eeq
where we have defined
\beq
z(\nu)=\int_{0}^{\nu}\frac{d\nu^\prime}{\epsilon_{\nu^\prime}}
\label{eq21}
\eeq
and used the fact that in $\overline
{f}_E$ appears $\overline
{z}(\nu)=z^*(\nu)$.
Evidently
\beq
\langle\overline{f}_E|f_{E^\prime}\rangle=|C|^2\int dz
\exp\left(\frac{i(E^\prime-E)z}{\alpha_sb}\right)=2\pi\alpha_sb|C|^2
\delta(E-E^\prime)\ , \label{eq22}
\eeq
and choosing $|C|=(2\pi \alpha_sb)^{-1/2}$
we obtain a correct normalization.

The most remarkable property of the solution is that it exists for
{\it any} real value of the energy $E$.
In the next Section we will show that there exists a normalizable
eigenfunction $\Psi_E(q)$ for any $-\infty < E < +\infty$.
Some doubts may arise in view of the
pole singularity  at $\nu=\nu_0$ generated by the denominator
$\epsilon^{(1)}(\nu)$.  However, inspection shows that at this point the
exponential provides a rapidly oscillating factor which ensures the convergence
of integrals involving
$f(\nu)$. So the spectrum
of the Hamiltonian, with the second order correction included,
extends
from
$-\infty$ to $+\infty$. Such a dramatic change in the spectrum is due to a
highly singular character of the ``interaction term" proportional to a $\ln
q$. From the mathematical point of view it is an operator unbounded from
below. With the rest of the Hamiltonian bounded from below, this shifts
the spectrum to arbitrary large negative values of $E$.
In the $\ln q$ space it is a linear potential which cannot be
considered as small perturbation irrespective of the magnitude of the
coupling constant which it accompanies. The solution (\ref{eq19}) is
accordingly
non-perturbative, the coupling constant appearing in the denominator of
the exponent.
There seems to be
no simple one-to-one
correspondence between the perturbative (i.e. 1st order)
and non-perturbative (i.e. 2nd order) wave functions.
A detailed discussion of the connection between perturbative and
non-perturbative solutions will be given in the next Section.


\section{Transition to the $q$ space. The high-energy limit}
In the $q$ space the found eigenfunctions are given by Eq. (\ref{eq11}).
We obtain
\beq
\Psi_E(q)=\frac{C}{\pi q\sqrt{2}}\int\frac{d\nu}{\epsilon_{\nu}}\exp
\left(i\nu\ln q^2-i\nu\frac{1}{\alpha_sb}+\frac{iEz(\nu)}{\alpha_sb}\right),
\label{eq23}
\eeq
with $z(\nu)$ defined by (\ref{eq21}).
Since $\alpha_s$ is supposed to be small we
can study the integral in $\nu$ by the saddle point method. The same method
will also provide us with the asymptotics of (\ref{eq23}) at very small and
very
large
$q$. The saddle point is determined by an equation
\[
Ez^\prime-1+\alpha_sb\ln q^2=0,
\]
which determines $\nu$ for a given eigenvalue $E$. It follows
\beq
E=\epsilon_{\nu}(1-\alpha_sb\ln q^2). \label{eq24}
\eeq
Putting this into (\ref{eq23}) one gets a crude asymptotic estimate
\beq
\Psi_E(q)\sim \frac{1}{q\epsilon_{\nu}}\exp
\left\{-i\left(\frac{1}{\alpha_sb}-\ln
q^2\right)[\nu-\epsilon_{\nu}z(\nu)]\right\}, \label{eq25}
\eeq
where $\nu$ should be determined from (\ref{eq24}).
Eq. (\ref{eq24}) is nothing but the usual expression for the pomeron energy $E$
as a function of $\nu$ with a running coupling constant to the
2nd order
$\alpha_s(q^2)=\alpha_s(1-\alpha_sb\ln q^2)$. However now we have to consider
it
as an equation for
$\nu$ at a given
$E$.

Let us discuss the behavior in $q$ of the solutions, for a {\it fixed} value
of $E$ (see Fig. 1). In all the arguments presented below, $\alpha_s$ is
small enough (smaller than
$\sim 0.05$, in order to avoid the complications found in
\cite{4r}) for the minimal value of $\epsilon_{\nu}$,
$\epsilon_{min}$,
to be\footnote{$\epsilon_{min}<0$; if we neglect the second
order
correction in $\epsilon_{\nu}$,
$\epsilon_{min}=-(\alpha_sN_c/\pi)\,4\ln 2$.} at $\nu=0$.

First we take $E>0$. For $\ln q^2<1/(\alpha_s b)$ (or $\alpha_s(q^2)>0$), 
(\ref{eq24}) has a pair of solutions
for real $\nu$ (Fig. 2), starting from $\pm \nu_0$ at
$\ln\, q^2 = -\infty$ and going to $\nu=\pm \infty$ when $\ln q^2$
approaches $1/(\alpha_s b)$.
According to (\ref{eq23}), for such
$q$ the eigenfunctions $\Psi_E(q)$ will essentially have a similar
behavior in
$q$ as the unperturbed eigenfunctions $\psi_{\nu}(q)$, with the correspondence
between $E$ and $\nu$ established by means of Eq. (\ref{eq24}): they are
plane waves in the variable $\ln q^2$,
apart from the common factor $1/q$ and a certain distortion due to the third
term in the exponent of (\ref{eq23})
(clearly visible in the form (\ref{eq25})). Thus we find an
one-to-one correspondence between the 1st order
perturbative and the 2nd order non-perturbative eigenfunctions
for this part of the $\ln q^2$ space; we call these oscillatory pieces of
the solutions ``normal".
With $\alpha_sb$ small this behavior remains valid up to very large values of
$q$ limited by the restriction $\ln q^2<1/(\alpha_sb)$.

What happens if $q$
becomes larger, so as $1/(\alpha_sb)<\ln q^2<(1-E/\epsilon_{min})/(\alpha_sb)$?
One can see that the picture changes and
Eq. (\ref{eq24})
will now give imaginary saddle points (Fig. 2),
starting from $\nu=\pm i/2$ for $\ln
q^2=1/(\alpha_sb)$ and approaching $\nu=0$ as $\ln q^2$ comes near
$(1-E/\epsilon_{min})/(\alpha_sb)$, so that the wave function
becomes damped  in $q$ by some power factors (as for the
case $E<0$, Eq. (\ref{eq25b}) but with an opposite sign in the exponent).
This piece of the eigenfunction which is governed by imaginary $\nu$ values
will be called
``abnormal".

Now we take $\ln q^2>(1-E/\epsilon_{min})/(\alpha_sb)$. Eq. (\ref{eq24})
has now solutions for real $\nu$, starting from $\nu=0$ and going to
$\pm \nu_0$ for $\ln q^2\rightarrow\infty$.
It is
easy to see the limiting asymptotics when $\ln q^2\rightarrow+\infty$.
Taking into account
the denominator $\epsilon_{\nu}$ in (\ref{eq23}) one then finds
an
oscillating
behavior\footnote{Actually
a sum of contributions from the
two points $\nu_0$, which differ only in sign if we neglect the 2nd
order
correction in $\epsilon_{\nu}$, should be taken.}
\beq
\Psi_E(q)\sim \frac{1}{q}\ \exp
\left(i\nu_0\ln q^2-i\frac{E}{c\alpha_s b}\ln\ln q^2
\right), \label{eq25a}
\eeq
where $c=\epsilon^\prime_{\nu_0}$.
This behavior is valid both for very small and
very large values of $q$ (and also for any sign of $E$, see below).

We consider now $E<0$ fixed in (\ref{eq24}).
Again we find three pieces in the solutions, corresponding to different
values of $\ln q^2$. For $\ln q^2<(1-E/\epsilon_{min})/(\alpha_sb)$,
(\ref{eq24}) will have solutions for real $\nu$, starting from $\pm \nu_0$
at $\ln q^2=-\infty$
and going to
$\pm \infty$ for $\ln q^2\rightarrow(1-E/\epsilon_{min})/(\alpha_sb)$.
This piece is equivalent to the last piece discussed in the case
$E>0$.

For $(1-E/\epsilon_{min})/(\alpha_sb)<\ln q^2<1/(\alpha_sb)$,
the corresponding values of the saddle point $\nu$ will have
a
non-zero imaginary part. If we neglect the second order correction to
$\epsilon_{\nu}$, Eq. (\ref{eq24}) will give a pair of pure
imaginary points $\pm i|\nu|$, with $|\nu|$ going from 0 at $\ln q^2=
(1-E/\epsilon_{min})/(\alpha_sb)$ to 1/2 for $\ln q^2= 1/(\alpha_sb)$.
To get rid of the singularities along the real axis it is convenient to
pass to the integration variable $z$ in (\ref{eq23}). Then, for $\ln
q^2$ close to $1/(\alpha_s b)$, in the complex $z$
plane the integrand will have singularities at points where
$1/\epsilon_{\nu}$ vanishes, that is, at poles of $\epsilon_{\nu}$.
They occur at pure imaginary $\nu=\pm i/2,\pm 3i/2,\dots$. Function $z$ at
these points will also take pure imaginary values $\pm i|z^{(k)}|$,
$k=1,2,\dots$, with $|z^{(1)}|<|z^{(2)}|<\dots$.
In the strip $|{\rm Im}\ z|<|z^{(1)}|$ the integrand will be analytic, so
that the integration contour in $z$ can be freely shifted up and down
parallel to the real axis. The solutions of Eq. (\ref{eq24}) for $E<0$ stay
inside this strip, tending to its boundaries as $E\rightarrow -\infty$
and correspondingly $\nu$ tends to $\pm i/2$.
So one can always shift the
integration contour to pass through the saddle point.
Note that if one takes into account the 2nd order correction to
$\epsilon_{\nu}$
then the first order poles at $\nu=\pm i/2$ are changed to third order poles
at the same point. As a consequence, the saddle points will acquire a real part
and tend to $\pm i/2$ at  certain angles when $E\rightarrow -\infty$, which
will however not influence the final result.
If we take $\alpha_s$ very small, with $q$ in this range, the
two saddle points approach their limiting values $\pm i/2$. The product
$\epsilon_{\nu}z(\nu)$ will be also pure imaginary with the same sign and its
modulus greater than $|\nu|$, since $|\epsilon_{\nu}|$ grows towards
$\nu=\pm i/2$. Then from (\ref{eq25})
it follows that we should shift the integration
contour down to pass through the saddle point with a negative imaginary part.
As a result we shall obtain
\beq
\Psi_E(q)\sim\frac{1}{q\epsilon_{\nu}}\exp
\left\{-\left(\frac{1}{\alpha_sb}-\ln
q^2\right)|\nu-\epsilon_{\nu}z(\nu)|\right\}, \label{eq25b}
\eeq
which shows that at finite $q$ the abnormal pieces are damped by the
non-perturbative damping factor $\exp (-\mbox{const}/\alpha_s)$. This
piece will correspond to the second piece discussed for $E>0$ (but, as
mentioned above, with
the opposite sign in the exponent: for $E<0$ $\ln q^2$ approaches
$1/(\alpha_s b)$ from below and for $E>0$ from above, see Fig. 1).

If we are now interested in the region $\ln q^2 >1/(\alpha_s b)$,
then
the saddle point starts from $\pm \infty$ at $\ln q^2 =1/(\alpha_s b)$
and
tends, for $\ln q^2 \longrightarrow +\infty$, to the point $\pm \nu_0$ at
which
$\epsilon_{\nu}=0$. As a result one obtains the same oscillating
asymptotic behavior as for the first and third pieces in $E>0$.

Excluding the region $\ln q^2>1/(\alpha_s b)$, which for small values of
$\alpha_s$ covers only exceptionally large values of $q$,
we can summarize the
situation saying that normal solutions (pieces)
have an oscillating behavior and
abnormal ones
are power damped both for small and large values of $q$ and also contain a
non-perturbative damping factor $\exp (-\mbox{const}
/\alpha_s)$. Abnormal solutions
have no correspondence with the perturbative ones. Since
there exist normalizable solutions
for arbitrary large negative values of energy, that is, for arbitrary
large
positive values of the angular momentum
$j$, they will lead to contributions to the amplitude which grow infinitely
fast
at high energies.

In fact even the passage to the energy representation results impossible
due to their existence.
The Green function of the total Hamiltonian as a function of
the angular momentum $j$ is given by the spectral representation
\beq G(j,q_1,q)=\int dE\frac{\Psi_E(q_1)\overline{\Psi}^*_E(q)}{j-1+E}
\ , \label{eq26}
\eeq
where the integration runs over all the spectrum. The Green function as
a function of rapidity $Y=\ln s$ is obtained by integrating (\ref{eq26}) with a
weight
$\exp{[Y(j-1)]}$,
\beq
G(Y,q_1,q)=\int dE\exp(-YE)\Psi_E(q_1)\overline{\Psi}^*_E(q).
\label{eq27}
\eeq
Evidently this integral is ill-defined for $Y>0$ if the integration goes
from arbitrary large negative values of $E$, as in our case.

However, throwing out non-perturbative contributions which behave
$\propto \exp (-1/\alpha_s)$, one obtains
an apparently reasonable asymptotics, which
coincides with the one found in \cite{2r}. Let us
cutoff the integral over $E$ in (\ref{eq27}) by some negative lower limit
$E_0<\epsilon_{min}$,
where $\epsilon_{min}$ is the discussed minimal value of energy $E$
for which Eq. (\ref{eq24}) gives real solutions for $\nu$:
\beq
G(Y,q_1,q)=\int_{E_0}^{\infty} dE\exp(-YE)\Psi_E(q_1)
\overline{\Psi}^*_E(q). \label{eq28}
\eeq

Putting our solutions we get then
\bea
G(Y,q_1,q)&=&\frac{|C|^2}{2\pi^2qq_1}\int_{E_0}^{\infty}dE\exp(-YE)
\int \frac{d\nu d\nu_1}{\epsilon_{\nu_1}}\nonumber \\
&&\exp [i\nu_1\ln q_1^2-i\nu\ln q^2-
i\beta(\nu_1-\nu)+i\beta E(z_1-z)], \label{eq29}
\eea
where we denoted $\beta=1/(\alpha_sb)$ for brevity. We integrate over $E$
to obtain
\bea
G(Y,q_1,q)&=&\frac{|C|^2}{2\pi^2qq_1}\exp(-YE_0)
\int \frac{d\nu d\nu_1}{\epsilon_{\nu_1}}\ [Y-i\beta(z_1-z)]^{-1}\nonumber \\
&&\exp [i\nu_1\ln q_1^2-i\nu\ln q^2-
i\beta(\nu_1-\nu)+i\beta E_0(z_1-z)]. \label{eq30}
\eea

Now we pass to the variable $z_1$. The integrand in (\ref{eq30}) has an
explicit
pole in it, and also singularities along the imaginary axis due to the
singularities of $\nu_1(z_1)$. The latter are situated at finite distance
from the real axis on both its sides. The explicit pole, on the contrary,
is quite close to the real axis, due to smallness of $\alpha_s$. It lies
slightly below the real axis in the $z_1$ plane. With $E_0<0$ we can close
the contour around the singularities in the lower $z_1$ semiplane.
The contribution from the cut along the negative semiaxis will contain
a damping factor $\exp (-1/\alpha_s)$. We shall neglect it, which is
certainly true in the limit $\alpha_s\rightarrow 0$ and amounts
to throwing out all anomalous solutions. Then we are left with only
 the residue at
\beq
Y-i\beta(z_1-z)=0. \label{eq31}
\eeq
In this approximation the dependence on $E_0$ disappears
and we get an integral over $\nu$:
\beq
G(Y,q_1,q)=\frac{|C|^2}{\pi qq_1\beta}
\int d\nu\exp [i\nu_1\ln q_1^2-i\nu\ln q^2-
i\beta(\nu_1-\nu)], \label{eq32}
\eeq
in which $\nu_1$ should be determined as a function of $\nu$ from  Eq.
(\ref{eq31}).
Solution of this equation can be accomplished by perturbation theory,
recalling that $\beta=1/(\alpha_sb)$ and is large.
Then we obtain in the first three orders
\beq
\Delta\nu=\nu_1-\nu=-\frac{iy}{z'}+\frac{y^2z''}{2(z')^3}-\frac{iy^3}{6(z')^5}
[z'z'''-3(z'')^2], \label{eq33}
\eeq
where we denoted $y=\alpha_sbY$.
To further simplify we take into account that the saddle point $\nu$ in
the integration  in (\ref{eq32}) is small, of order $1/Y$. This
allows to use an approximation
\beq
\epsilon_{\nu}=\epsilon_0+i\delta\nu+\overline{a}\nu^2, \label{eq34}
\eeq
with \cite{1r}
$$\epsilon_0=-\frac{N_c\alpha_s}{\pi}
\ 4\ln{2}\left[1-\frac{N_c\alpha_s}{4\pi}\left(25.8387+0.1869
\ \frac{N_f}{N_c}+
3.8442\ \frac{N_f}{N_c^3}\right)\right],$$
$$\delta=\left(\frac{N_c\alpha_s}{\pi}\right)^2\left(15.4262-2.8048
\ \frac{N_f}{N_c}\right),$$
$$\overline{a}=a-\left(\frac{N_c\alpha_s}{\pi}\right)^2\left
(322.188-3.10189\ \frac{N_f}{N_c}+21.6732
\ \frac{N_f}{N_c^3}\right),\ \
a=\frac{N_c\alpha_s}{\pi}\ 14\zeta(3) $$
($\zeta(z)$ being
Riemann's zeta function),
and to restrict ourselves to terms up to the second order in $\nu$ in the
exponent in (\ref{eq32})\footnote{The
values of $\alpha_s$ considered here are those
small enough to keep $\overline{a}>0$, i.e. $\alpha_s$ smaller than
$\sim 0.05$. The case $\overline{a}<0$ has been examined in
\protect{\cite{4r,levin}} and leads to an oscillatory behavior of the
Green function.}. The exponent then becomes
a polynomial
\beq
P(\nu)=p_0-ip_1\nu-p_2\nu^2, \label{eq35}
\eeq
where from (\ref{eq33}) and
(\ref{eq34}) we find (up to terms $\propto \alpha_s^{n+2}
Y^n$)
\bea
p_0&=&-Y\epsilon_0(1-\alpha_sb\ln
q_1^2)+\frac{1}{3}(\alpha_sb\epsilon_0)^2aY^3-\frac{1}{2}\delta
\alpha_sb\epsilon_0 Y^2,
\nonumber \\
p_1&=&\ln\frac{q^2}{q_1^2}-\alpha_sba\epsilon_0 Y^2+\delta Y,
\nonumber \\
p_2&=&aY. \label{eq36}
\eea
Integration over $\nu$ gives the desired asymptotics at large $Y$:
\beq
G(Y,q_1,q)=\frac{|C|^2}{\pi qq_1\beta}\sqrt{\frac{\pi}{p_2}}
\exp \left(p_0-\frac{p_1^2}{4p_2}\right). \label{eq37}
\eeq
Putting expressions (\ref{eq36}) into (\ref{eq37}), retaining only
terms up to $Y^3$ in the exponent and neglecting terms of order
$(\delta/a)$ we obtain
\bea
G(Y,q_1,q)&=&\frac{|C|^2}{\pi qq_1\beta}\sqrt{\frac{\pi}{aY}}
\label{eq38}
\\
&&\exp\left (-Y\epsilon_0\left[1-\frac{\alpha_s b}{2}(\ln q^2+\ln q_1^2)
\right]-
\frac{\ln^2(q^2/q_1^2)}{4aY}+\frac{1}{12}(\alpha_sb\epsilon_0)^2aY^3
\right).\nonumber
\eea
This expression coincides with the asymptotics found in \cite{2r}, if we take
into account that our $a=4D$ in \cite{2r}.

\section{Running coupling in all orders}
An amusing application of the described formalism is a possibility
to write an infrared stable equation
for the case when  the interaction is taken only
to the first order in the coupling, but the coupling is taken running in all
orders. In other words
\beq
H\psi_\nu=\epsilon^{(1)}_{\nu}(q)\psi_{\nu}(q), \label{eq39}
\eeq
where
\beq
\epsilon^{(1)}_{\nu}(q)=\frac{N_c}{\pi b\ln\frac{q^2}{\Lambda^2}}
\epsilon^{(1)}_{\nu} \label{eq40}
\eeq
and $\epsilon^{(1)}_{\nu}$ is defined by (\ref{eq5}). In the following we put
$\Lambda=1$.

Of course, we understand that
(\ref{eq40}) is quite unsatisfactory from the physical
point of view: it implies that the coupling constant is prolonged as a pure
imaginary quantity into the confinement region $q<1$. So the following serves
only as an illustration of the power of the employed technique and also as a
warning against a simple-minded use of the saddle point approximations.
Note that in \cite{levin},
where, as mentioned in the Introduction, a similar problem is
solved by  a similar technique, a different choice was made:
\[ \ln q^2\ \rightarrow\ r(q), \]
where $r(q)$ was to be determined from a transcendental equation
\[ r=\ln q^2+\frac{1}{2}\ \ln (br). \]
However for $q^2<\sqrt{2e/b}$ this equation leads to complex values of $r$ and
consequently for $\alpha_s(q^2)=1/(br)$, which seems to be even worse than
our choice.

In the $\nu$ representation we obtain the Hamiltonian as
\beq
H(\nu,\nu')=\frac{N_c}{2\pi b}i\epsilon^{(1)}_{\nu'}{\rm sign}(\nu'-\nu).
\label{eq41}
\eeq
The eigenvalue equation becomes
\beq
\frac{N_c}{2\pi b}i\int d\nu'\epsilon^{(1)}_{\nu'}{\rm sign}(\nu'-\nu)
f(\nu')=Ef(\nu). \label{eq42}
\eeq
Differentiating with respect to $\nu$ we obtain
\beq
-\frac{N_c}{\pi b}i\epsilon^{(1)}_{\nu}f(\nu)=Ef'(\nu), \label{eq43}
\eeq
with a solution
\beq
f(\nu)=\frac{C}{E}
\exp\left(-i\frac{N_c}{\pi bE}\int_{0}^{\nu}d\nu'\epsilon^{(1)}_{\nu'}
\right). \label{eq44}
\eeq
This seems to be a valid solution for any real $E$, positive or negative.
The conjugate equation is
\beq
-\frac{N_c}{\pi b}i
\overline{f}(\nu)=E\left(\overline{f}(\nu)/\epsilon^{(1)}_{\nu}\right)',
\label{eq45}
\eeq
with  solutions similar to (\ref{eq44}) but with an extra factor
$\epsilon^{(1)}_{\nu}$.
They also exists for any $E$. Requiring
$\langle\overline{f}_E|f_{E'}\rangle=\delta(E-E')$, we get
$|C|=(2\pi^2b/N_c)^{-1/2}$. Solutions (\ref{eq44}) and its conjugate can be
readily expressed in terms of the $\Gamma$-function. Indeed
\beq
z(\nu)=\int_0^{\nu}d\nu'\epsilon^{(1)}_{\nu}=-i\ln\frac{\Gamma
(1/2+i\nu)}
{\Gamma(1/2-i\nu)}-2\nu\ \psi(1). \label{eq46}
\eeq

In the momentum representation the eigenfunctions have the form
\beq \Psi_E(q)=\frac{C}{\pi E q\sqrt{2}}\int d\nu \exp\left(
 i\nu\ln q^2- i\frac{N_c}{2\pi bE}z \right). \label{eq47}
\eeq
 To find the asymptotics at
large $|\ln q^2|$ we employ the saddle point approximation.  The
saddle point is determined by a simple equation
\[
\ln q^2=\frac{N_c}{\pi bE}\epsilon^{(1)}_{\nu}
\]
or \beq E=\frac{N_c}{\pi b\ln
q^2}\epsilon^{(1)}_{\nu}, \label{eq48}
\eeq
where we used the definition (\ref{eq40}). This equation is again nothing but
the relation between the energy and $\nu$, expected from (\ref{eq39}).
It has to
be
considered as an equation for the saddle point $\nu$ for a given $E$.
We will now analyze, as in the previous Section, the behavior of the
solutions in momentum space, for fixed $E$ (having in mind that the
saddle point will only give a reliable solution for very large $|\ln
q^2|$).

Let us take $E>0$. Then, for $\ln q^2< N_c\epsilon_{min}^{(1)}/(\pi
bE)=-4N_c\ln 2/(\pi bE)$,
(\ref{eq48}) has solutions for imaginary $\nu$,
going from
$\nu=\pm i/2$ at $\ln q^2=-\infty$ to $\nu=0$ at $\ln
q^2=N_c\epsilon_{min}^{(1)}/(\pi
bE)$. As a result, for (\ref{eq47}) we obtain a falling asymptotics (see
below the case $E<0$); this would correspond, in the
language of the previous Section, to an abnormal piece.

Now, if $\ln q^2>N_c\epsilon_{min}^{(1)}/(\pi
bE)$, (\ref{eq48}) has solutions for real $\nu$, going from
$\nu=0$ at $\ln
q^2=N_c\epsilon_{min}^{(1)}/(\pi
bE)$ to $\nu=\pm \infty$ for $\ln q^2=+\infty$.
This will result in an oscillating
asymptotics for $\Psi_E(q)$ (the normal piece),
apart from the dimensionful factor $1/q$.

We consider now the case $E<0$. For $\ln q^2<N_c\epsilon_{min}^{(1)}/(\pi
bE)$, the saddle point equation (\ref{eq48}) will give real solutions,
going from $\nu=\pm \infty$ for $\ln q^2=-\infty$ to $\nu=0$ at $\ln
q^2=N_c\epsilon_{min}^{(1)}/(\pi
bE)$. The solution (\ref{eq47}) will be, as for the second piece in the
case $E>0$, a normal oscillating piece.

For $\ln q^2>N_c\epsilon_{min}^{(1)}/(\pi
bE)$,
(\ref{eq48}) will have a pair of conjugate purely
imaginary solutions for $\nu$. At large $\ln q^2$ they approach points
$\pm i/2$. In the vicinity of, say, $\nu=i/2$ we have
\[ \epsilon^{(1)}_{\nu}\simeq-(1/2+i\nu)^{-1}, \]
so that from (\ref{eq48}) we find
\beq
\frac{1}{2}+i\nu=-\frac{N_c}{\pi b E\ln q^2}\ . \label{eq49}
\eeq
Function $z$ has a logarithmic singularity at this point so that it
will be approximately given by
\beq
z\simeq i\ln{\left|\frac{1}{2}+i\nu\right|}=i\ln\frac{N_c}{\pi b|E|\ln q^2}
\ , \label{eq50}
\eeq
where we have taken into account that in the considered region $E<0$.
It follows that at this saddle point the exponent in (\ref{eq47}) becomes
\[  i\nu\ln q^2- i\frac{N_c}{\pi bE}z\simeq-\ln q^2\left[
\frac{1}{2}-\frac{N_c}{\pi b|E|\ln{q^2}}\left(1-
\ln{\frac{N_c}{\pi b|E|\ln{q^2}}}\right)\right].\]
At the complex conjugate saddle point, in the vicinity of $\nu=-(1/2)i$,
the exponent will have an opposite sign.
In the complex $\nu$ plane the integrand in (\ref{eq47}) has singularities
along
the
imaginary axis for $|{\rm Im}\ \nu|>1/2$. In the  strip $|{\rm
Im}\ \nu|<1/2$
it is analytic, so that the integration contour can be freely shifted
above
or below the real axis in this interval. Depending on the sign of
\[\frac{1}{2}-\frac{N_c}{\pi b|E|\ln{q^2}}\left(1-
\ln{\frac{N_c}{\pi b|E|\ln{q^2}}}\right),\]
one can always shift the contour to pass over one of the two conjugate
saddle points so as to have a negative coefficient before $\ln q^2$. As
a
result we obtain a falling asymptotic (abnormal)
behavior in this $\ln q^2$ range
(corresponding to the first piece in the case $E>0$):
\beq
\Psi_E(q)\propto \frac{1}{q}\ q^{-\left|1-\frac{2N_c}{\pi b |E|\ln{q^2}}
\left(1-
\ln{\frac{N_c}{\pi b|E|\ln{q^2}}}\right)\right|}. \label{eq51}
\eeq

Summarizing, both for $E>0$ or $E<0$ we find that the solution has two
pieces, normal and abnormal, depending on the $\ln q^2$ range we are
studying. The eigenfunction (\ref{eq47}) is normalizable for any real value
of $E$. Restricting ourselves to the case $\ln
q^2>N_c\epsilon_{min}^{(1)}/(\pi
bE)$ (corresponding to $\ln q^2 > 0$ for large negative $E$),
we find a normal oscillating solution for $E>0$ and an abnormal falling
solution for $E<0$.

With the spectrum of the Hamiltonian (\ref{eq39}) extending from $-\infty$ to
$\infty$,
we again have grave problems in passing from the angular momentum
to rapidities. From the start the integral (\ref{eq27}), which determines the
Green
function as a function of rapidity, is badly divergent at large negative
$E$.
It is amusing that nevertheless one obtains  a seemingly reasonable
asymptotics at high $Y$ and momenta if one forgets about the initial
divergence and makes some crude approximations. Indeed, with the
eigenfunctions given by (\ref{eq47}), we have
\bea
G(Y,q_1,q)&=&\frac{|C|^2}{2\pi^2qq_1}\int\frac{dE}{E^2}\ \exp(-YE)
\int d\nu_1 d\nu\ \epsilon_{\nu}^{(1)}\nonumber \\
&&\exp \left(i\nu_1\ln q_1^2-i\nu\ln q^2
-i\frac{N_c}{\pi bE} (z_1-z)\right), \label{eq52}
\eea
where $z_1=z(\nu_1)$. The integral (\ref{eq52})  does not exist.
However let us forget this and calculate its asymptotics for large $\ln
q^2$
 and
$\ln q_1^2$. Then  small values of $\nu$ and $\nu_1$ and
consequently of $z$ and $z_1$ evidently give the dominant contribution.
At small $\nu$
\[z=\epsilon^{(1)}_0\nu,\]
so that we obtain
\[
G(Y,q_1,q)=\frac{2|C|^2\epsilon^{(1)}_0}{qq_1}\int\frac{dE}{E^2}\ \exp(-YE)
\ \ \delta\left(\ln q^2-\frac{N_c\epsilon^{(1)}_0}{\pi bE}\right)
\ \ \delta\left(\ln q_1^2-\frac{N_c\epsilon^{(1)}_0}{\pi bE}\right).\]
This integral exists, since the $\delta$-functions ensure that the
integrand
is zero nearly for all $E$ and for large negative $E$ in particular.
 We obtain in this manner
\beq
G(Y,q_1,q)=\frac{|C|^2\pi b}{N_c}\ \frac{\delta(q-q_1)}{q}
\exp\left(-Y\frac{N_c\epsilon^{(1)}}{\pi b\ln q^2}\right). \label{eq53}
\eeq
The behavior in $Y$ is just what one would expect from naive
expectations
of having a running coupling in the intercept.

Of course, this exercise has very little meaning. It only shows that using
poorly controlled approximations one can arrive at seemingly
sensible results even though the exact expression has no meaning at all.
In our case it shows that the gluon interaction should have a different
functional dependence to cure its bad qualities, which arise when one
simply changes a fixed coupling by a running one. In mathematical
terms the interaction then results not bounded from below, which shifts
the spectrum down to $-\infty$.

\section{Conclusions}

Different aspects of the 2nd order corrections to the
BFKL pomeron \cite{1r,3r} have been studied recently \cite{4r,2r,levin,bv}
using different techniques. Some of the resulting features, as the
existence of a ``non-Regge" term $\propto \exp{(Y^3)}$ \cite{2r,levin} or
of oscillatory solutions \cite{4r,levin} for $\alpha_s$ greater than $\sim
0.05$, together with the diffusion into the infrared region
\cite{2r,mueller}, impose severe limits on the possible applicability of
these results to the experimental situation.

In this note we find the solutions to the 2nd order BFKL Hamiltonian and
analyze the resulting spectrum
in the energy plane ($E=1-j=-\omega$).
It turns out not to be bounded from below, and we show that for any
$-\infty < E < +\infty$ there exists a normalizable eigenfunction
$\Psi_E(q)$. This solution exhibits non-perturbative features,
with the coupling constant
appearing in the denominator of the exponent, Eq. (\ref{eq23}).
Both for the cases $E>0$ and $E<0$, solutions to the eigenvalue
equation
have three different
pieces in momentum
space, depending on the relation of the value of $\ln q^2$ with
$\alpha_s$ and $E$. The different pieces can be classified as being ``normal",
with an oscillating behavior as in the 1st order case, and ``abnormal",
which are non-oscillatory
in $q$ and have no equivalence to the 1st order solutions.
For very small $\alpha_s$, the case $E>0$ is almost entirely normal,
whereas for $E<0$ the eigenfunction is almost everywhere abnormal.
From our study we conclude that, in the presence of
the 2nd order corrections,
the unbounded spectrum of the kernel makes the use of the complex
angular momentum formalism to study the high-energy behavior somewhat
doubtful.

A calculation of the high energy behavior of the 2nd order BFKL Green
function has been presented in \cite{2r}. In this paper,
Mellin transforms were applied only to the leading-log BFKL Green
function, and the problem of the unbounded spectrum of the second order
corrections was avoided.
So the iteration techniques used in \cite{2r} seem to offer a
possibility to compute the high energy behavior
of the 2nd order BFKL Green function bypassing the
difficulties of the unbounded spectrum found in our paper. We have shown
that with a modification of the Mellin transform the Green function can
be defined by simply excluding the abnormal pieces of the
eigenfunctions; this definition leads to a high energy behavior which
agrees with \cite{2r}.

In \cite{levin} also the high-energy behavior of the 2nd order BFKL Green 
function is
investigated. The result is consistent with Ref. \cite{2r}. Although the
methods used in \cite{levin} are similar to parts of our study, no
particular attention has been paid to the energy spectrum of the 2nd order
BFKL kernel which constitutes the main purpose of our paper.

So far we have concentrated on the spectrum of the 2nd order BFKL kernel,
as it was presented in ~\cite{1r}. In the context of the leading-order
BFKL equation ~\cite{Lip,cc} it has already been emphasized that the 
physical spectrum of the Pomeron depends upon the (nonperturbative) infrared
behaviour, i.e. upon the (unknown) extrapolation of the perturbative BFKL
kernel into the small-$q$ region. We have not yet adressed this question,
but we feel that the arguments which in ~\cite{Lip,cc} have been given for the
leading-order BFKL equation should now be reconsidered in presence of the
2nd order corrections which lead to this dramatic change of the energy 
spectrum.

\vskip 1 cm
\noi {\bf Acknowledgments:}
The authors express their gratitude to Profs. L.N.Lipatov and E.Levin 
for helpful comments on the
manuscript. M. A. B. is thankful to the Universit\"at of
Hamburg for the financial support and to DESY for the hospitality during his 
stay in Hamburg where this paper was initiated. N. A. gratefully acknowledges
the financial support from Direcci\'on General de Investigaci\'on 
Cient\'{\i}fica 
y T\'ecnica of Spain and the hospitality of the II.Institut f.Theor.Physics 
of the University of Hamburg.

\vskip 3cm

\begin{figure}[htb]
\hspace{1.5cm}\epsfig{file=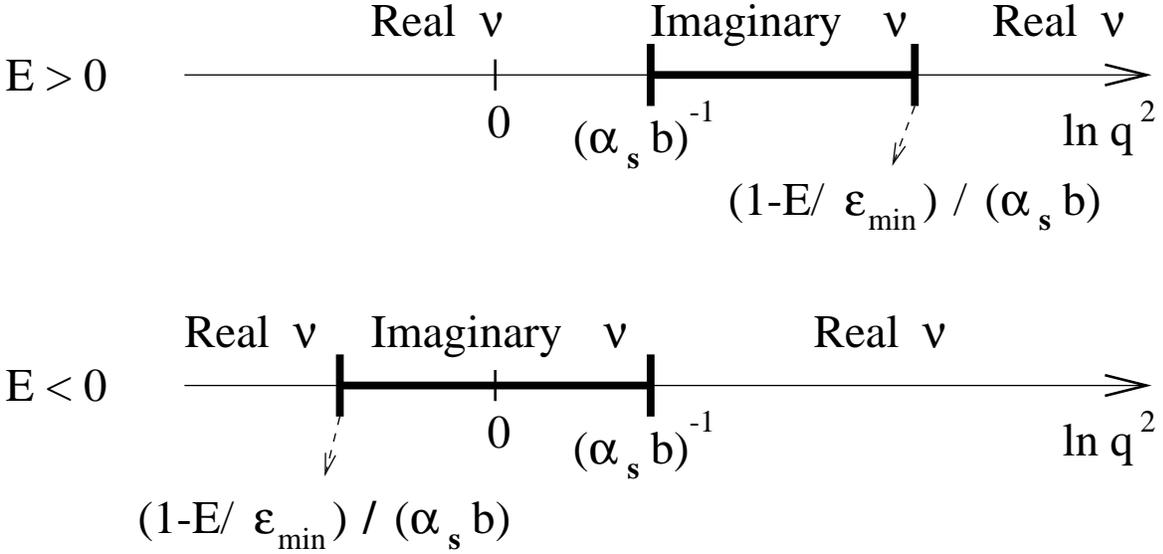,width=7.4cm,angle=270}
\caption{Behavior in $q$ of the solutions (\protect{\ref{eq23}}) for fixed
$E>0$ and $E<0$. Real $\nu$ correspond to an oscillatory (normal)
piece, while
imaginary $\nu$ correspond to an exponentially damped (abnormal) one.}
\end{figure}

\vskip 3cm

\begin{figure}[htb]
\hspace{1.5cm}\epsfig{file=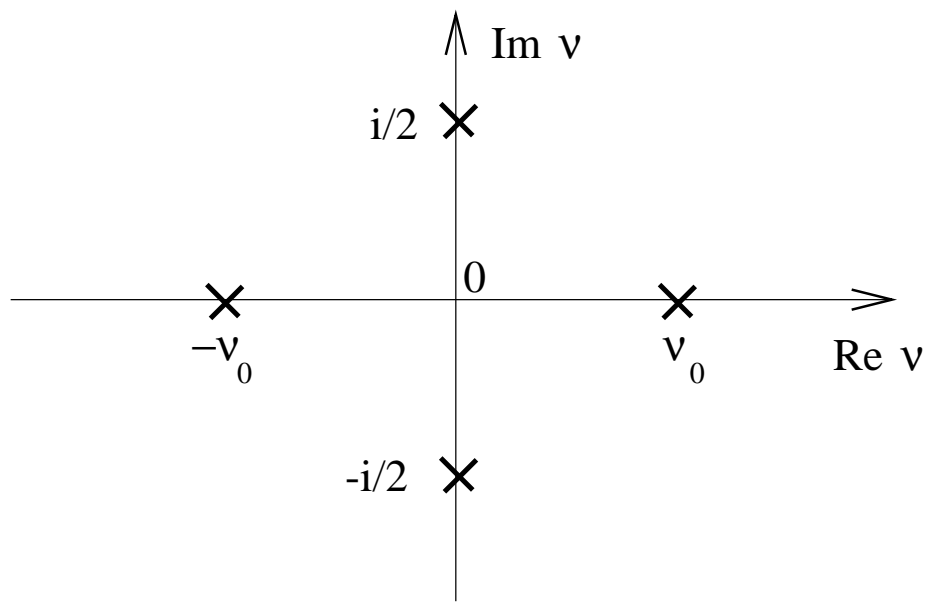,width=8cm,angle=270}
\caption{Location of zeros ($\nu=\pm \nu_0$) and poles ($\nu=\pm i/2$) of
$\epsilon_\nu$ in the complex $\nu$ plane.}
\end{figure}
%
%

\end{document}